\documentclass[fleqn,a4paper]{article}

\usepackage{graphicx}

\begin{document}
\raggedright
\sffamily

\title{On the propagation of neutrinos through the Earth}
\author{M.\ de Jong\\
  {\small NWO-I, Nikhef, PO Box 41882, Amsterdam, 1098 DB Netherlands} \\
  {\small Leiden University, Leiden Institute of Physics, PO Box 9504, Leiden, 2300 RA Netherlands}
}
\maketitle

\begin{abstract}
The Earth is commonly used as a natural filter for the operation of deep-underground and deep-sea neutrino telescopes.
By selecting events pointing in upward directions, 
the background of muons produced by interactions of cosmic rays in the Earth' atmosphere above the detector can effectively be suppressed.
The surviving neutrinos traversed a large part of the Earth before being detected.
It is commonly assumed that the detected neutrinos go in a straight line through the Earth without loosing energy.
A first study has been made of the propagation of neutrinos through the Earth which includes 
the effects of the charged-current as well as neutral-current interactions.
It is found that this leads to an increase of the detectable flux of neutrinos.
\end{abstract}

\section{Introduction}

The cross-section of neutrinos to interact with normal matter is known to be very small compared to that of other elementary particles.
This makes it relatively difficult to detect neutrinos.
On the other hand, it is possible to filter neutrinos by employing a sufficiently thick layer of matter.
For the operation of neutrino telescopes such as IceCube and KM3NeT, 
the Earth is used a a natural filter for the detection of high-energy neutrinos ($1~\mathrm{TeV} - 1~\mathrm{PeV}$) 
from the cosmos \cite{ref:icecube,ref:km3net}.
By selecting events pointing in upward directions (i.e.\ by looking downward), 
the background of muons produced by interactions of cosmic rays in the Earth' atmosphere above the detector can effectively be suppressed.
The corresponding neutrinos traversed a large part of the Earth before being detected.
For the interpretation of the data, 
it is thus important to know how a flux of neutrinos is affected by the propagation of neutrinos through the Earth.
In the following, the cross-section of neutrinos to interact with matter is taken from reference \cite{ref:cross-section}.
Up to energies of $1~\mathrm{TeV}$ or so, the cross-section is so small 
that (almost) all neutrinos pass through the Earth unhampered.
As a function of energy, however, the cross-section steadily increases. 
As a consequence, (almost) all neutrinos interact on their way through the Earth at energies in excess of a few $\mathrm{PeV}$.
The interaction of a neutrino in the Earth is usually interpreted as absorption, 
thereby leaving a finite energy window to observe neutrinos from the other side of the Earth.
This limitation can partly be overcome by also selecting horizontally or even downward traveling neutrinos 
which comes with the cost of a (much) larger background.\\[\baselineskip]

\begin{figure}[h]
\begin{center}
\begin{picture}(14,5)
\put(1,0){\includegraphics[trim={13cm 6cm 13cm 6cm},clip,height=5cm]{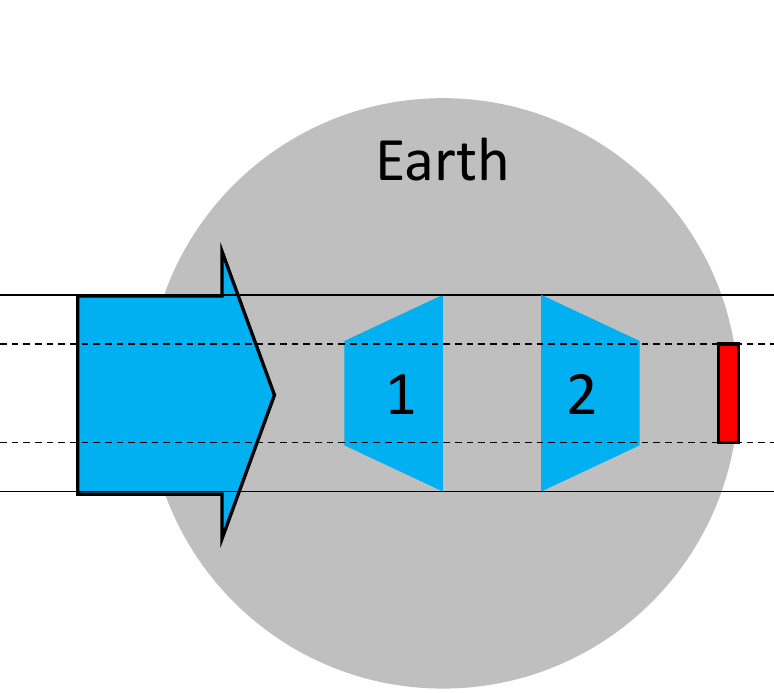}}
\put(8,0){\includegraphics[trim={14cm 6cm 13cm 6cm},clip,height=5cm]{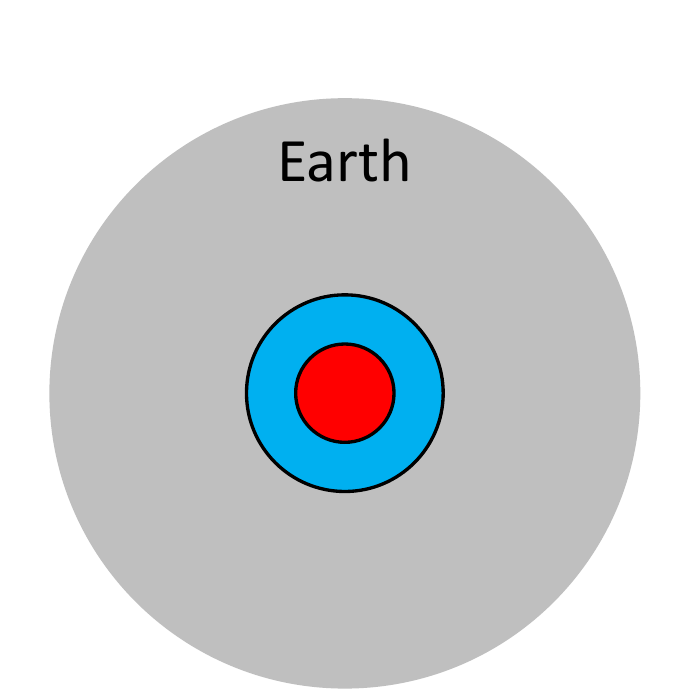}}
\end{picture}
\end{center}
\caption{\label{fig:setup}
Schematic views of the propagation of neutrinos through the Earth (not to scale).
On the left, side view and on the right, front view.
A parallel flux of neutrinos (indicated by the blue arrow and blue disk, respectively) is pointing to the detector (indicated by red area).
Due to scattering, 
part of the flux which overlaps with the detector may miss the detector (indicated by divergent blue area labeled 1) and
part of the flux which extends beyond the detector may hit the detector (indicated by convergent blue area labeled 2).
}
\end{figure}

Neutrinos can interact with matter via the so-called charged-current or neutral-current interaction 
in which a $W^{\pm}$ or $Z$ boson is exchanged with a nucleon in the Earth, respectively.
In the first, the neutrino is transformed into a corresponding charged lepton.
With the exception of the $\tau$-lepton, 
the charged lepton usually stops in the Earth before reaching the detector 
without leaving any trace of the original neutrino \cite{ref:halzen-saltzberg}.
In the second, the neutrino looses energy and changes direction but continues thereafter.
As a consequence, a flux of neutrinos will be attenuated and diffused.
In the following, the effect of diffusion on the detectable flux in a neutrino telescope is investigated.
To this end, 
a Mickey-mouse simulation has been made of the propagation of neutrinos through the Earth 
which includes attenuation as well as diffusion.
Without loss of generality, only neutrinos propagating through the complete Earth are considered.
A schematic view of the set-up is shown in figure \ref{fig:setup}.
In this, a homogeneous flux of parallel neutrinos is entering the Earth from the opposite side of a hypothesized detector.\\[\baselineskip]

For the propagation of the neutrinos through the Earth,
charged-current as well as neutral-current interactions are considered.
The first directly leads to absorption whereas the second leads to a loss of energy and a change of direction.
A rudimentary simulation of the scattering angle distribution is made by assuming 
an isotropic distribution in the center-of-mass of the neutrino and the struck nucleon.
This angular distribution is motivated by the large mass of the $W$ and $Z$ boson.
Here, the main point is that even a small scattering angle can cause a large deviation 
because the diameter of the Earth is generally way larger than the size of the detector.
The energy loss for neutrinos and anti-neutrinos follows from 
the corresponding Bj{\"o}rken-y distributions and 
the (anti-)quark structure of the nucleons.
The diameter and average density of the Earth are taken to be $12724~\mathrm{km}$ and $5.5~\mathrm{g/cm^3}$, respectively.

\section{Results}

The "survival probability" is determined from the number of neutrinos passing through the detector due to a given flux.
A sample of 1000,000,000 muon neutrinos with a fixed energy is generated on 
the opposite side of the Earth (referred to as source) and 
subsequently propagated through the Earth towards the detector (referred to as target).
In this, the number of neutrinos and anti-neutrinos relate $1:1$.
The assumed target is a circular area with a radius of $500~\mathrm{m}$.
This roughly corresponds to the size of the IceCube and KM3NeT neutrino telescopes.
At the source, a circular area perpendicular to the flux is randomly covered. 
The numbers of neutrinos, $N$, which hit the target are separately evaluated 
for neutrinos that started at the source in an area which is either 
\begin{enumerate}
\item covered by the target, referred to as ``inside''; or 
\item not covered by the target, referred to as ``outside''.
\end{enumerate}

A comparison is then made between 
a realistic scenario in which charged-current and neutral-current interactions are jointly simulated and 
a conservative scenario in which any interaction is attributed to absorption.
In the latter scenario, the second number is zero by definition.
The ratio of the corresponding survival probabilities is then equal to $R = (\sum\: N_{A}) / N_{B}$, 
where $A$ and $B$ refer to the realistic and conservative scenario, respectively.
Before that, the radius of the source area was tuned as follows.
The radius of the target was taken as a start value.
It was subsequently enlarged up to the point where the ratio of survival probabilities no longer increases.
The thus obtained radius of the source area is 80 times larger than that of the target.
This large value can be explained by the diameter of the Earth and the typical scattering angle.
Due to energy and momentum conservation, this angle roughly amounts to $\sqrt{2M_{N}/E_{\nu}}$, 
where $M_{N}$ is the average nucleon mass and $E_{\nu}$ the neutrino energy.\\[\baselineskip]

The numbers of neutrinos which hit the target are listed as a function of the neutrino energy in table \ref{tab:results}.
In this, the columns $C$, $D$ and $E$ refer to the hypothetical scenarios in which 
there is no energy loss, 
no scattering no or
no interaction at all due to neutral-currents, respectively.
In scenarios $D$ and $E$, the number of neutrinos from outside is zero by definition.
By comparing the values in columns inside and outside of column $A$, 
it can be seen that the contribution of neutrinos from outside can be much larger than that from inside. 
This can be explained by the effective source area which is much larger than the target area. 
By comparing the values in column $C$ to those in column $A$, 
it can be seen that the contribution of neutrinos from outside is smaller when the energy loss in neutral-current interactions is not taken into account.
This can be understood as follows.
Due to the energy loss in a neutral-current interaction, 
the cross-sections of a neutrino to interact again is reduced.
As a consequence, the probability to be absorbed during the remaining path is reduced.
This reduction applies to all neutrinos but it is correlated with the scattering of the neutrinos required to hit the target from outside.
It is interesting to note that also in the absence of energy loss, 
the number of neutrinos from outside can exceed 
the number of neutrinos from inside.\\[\baselineskip]

As can be seen from the table \ref{tab:results},
the ratio of survival probabilities (column $R$) is larger than one for neutrino energies in excess of few $\mathrm{TeV}$.
This could be interpreted as a realistic gain of the detectable flux of neutrinos when comparing to a conservative scenario.
Compared to the hypothetical scenario in which there are no interactions at all due to neutral-currents (column $E$),
the number of neutrinos hitting the target is also larger but then for neutrino energies in excess of $20~\mathrm{TeV}$.
This could also be interpreted as a realistic gain of the detectable flux of neutrinos when comparing to a simplified scenario.
It should be noted, however, that the energy of the surviving neutrinos generally is overestimated in the simplified scenario.
The match between the numbers of neutrinos hitting the target 
in the realistic scenario (column $A$) and 
in the hypothetical scenario in which there is no scattering due to the neutral-current interactions (column $D$) 
could provide for practical solutions avoiding the use of source areas which are (much) larger than the target area.\clearpage

\begin{table}[ht]
  \begin{center}
    \begin{tabular}{|r|r@{\,}r|r|r@{.}l|r@{\,}r|r|r|}

      \hline
      \multicolumn{1}{|c|}{$E_{\nu}$}        &
      \multicolumn{2}{c|}{$A$}              &
      \multicolumn{1}{c|}{$B$}              &
      \multicolumn{2}{c|}{$R$}              &
      \multicolumn{2}{c|}{$C$}              &
      \multicolumn{1}{c|}{$D$}              &
      \multicolumn{1}{c|}{$E$}              \\

      \multicolumn{1}{|c|}{[GeV]}           &
      \multicolumn{1}{c}{inside}            &
      \multicolumn{1}{c|}{outside}          &
      \multicolumn{1}{c|}{}                 &
      \multicolumn{2}{c|}{}                 &
      \multicolumn{1}{c}{inside}            &
      \multicolumn{1}{c|}{outside}          &
      \multicolumn{1}{c|}{}                 &
      \multicolumn{1}{c|}{}                 \\
      \hline
      1.0e+03 &  152,220 &      430 &  152,207 &          1&00         &  152,159 &      435 &  153,204 &  153,148  \\
      2.0e+03 &  148,455 &    1,000 &  148,418 &          1&01         &  148,484 &      986 &  150,396 &  150,409  \\
      5.0e+03 &  139,190 &    2,982 &  139,091 &          1&02         &  139,241 &    2,936 &  143,409 &  143,401  \\
      1.0e+04 &  125,818 &    5,701 &  125,612 &          1&05         &  125,698 &    5,555 &  132,922 &  132,580  \\
      2.0e+04 &  107,884 &   10,119 &  107,493 &          1&1          &  107,954 &    9,581 &  118,757 &  118,290  \\
      5.0e+04 &   77,481 &   16,007 &   76,569 &          1&2          &   77,630 &   14,545 &   93,852 &   92,350  \\
      1.0e+05 &   52,249 &   18,868 &   50,977 &          1&4          &   52,115 &   16,315 &   70,928 &   68,236  \\
      2.0e+05 &   30,390 &   18,986 &   28,859 &          1&7          &   30,142 &   15,136 &   49,195 &   45,019  \\
      5.0e+05 &   10,981 &   14,644 &    9,621 &          2&7          &   10,797 &    9,504 &   25,479 &   20,331  \\
      1.0e+06 &    4,032 &    9,867 &    2,983 &          4&7          &    3,733 &    4,965 &   13,861 &    8,763  \\
      2.0e+06 &    1,214 &    5,565 &      566 &         12&0          &      921 &    1,730 &    6,782 &    2,714  \\
      5.0e+06 &      313 &    2,473 &       35 &         80&0          &      106 &      223 &    2,745 &      335  \\
      1.0e+07 &      182 &    1,302 &        0 & \multicolumn{2}{c|}{} &       18 &       30 &    1,465 &       31  \\
      \hline
    \end{tabular}
  \end{center}
\caption{\label{tab:results}
Number of neutrinos which hit the target as a function of the neutrino energy at the source.
The column $A$ corresponds to the realistic scenario in which charged-current and neutral-current interactions are jointly simulated and
the column $B$ to the conservative scenario in which any interaction is attributed to absorption.
The ratio of survival probabilities is then equal to $R = (\sum\: N_{A}) / N_{B}$.
The columns $C$, $D$ and $E$ correspond to the hypothetical scenarios in which 
there is no energy loss, 
no scattering no or
no interaction at all due to neutral-currents, respectively.
The columns inside and outside correspond to neutrinos that started at the source in an area which is covered or not by the target.
}
\end{table}

The main systematic uncertainty of these results is related to the interaction length of neutrinos 
$\lambda^{-1} = \sigma \: N_{A} \: \rho$,
where 
$\sigma$ corresponds to the cross-section of neutrinos to interact with matter,
$N_{A}$ to Avogadro's number and
$\rho$ to the density of the Earth.
As stated before, the cross-section depends on the energy of the neutrino (for an overview, see e.g.\ reference \cite{ref:eVEV}). 
The density of the Earth varies considerably, 
between less than $2.7~\mathrm{g/cm^3}$ in the upper crust to 
as much as $13~\mathrm{g/cm^3}$ in the inner core.
The (local) ratio of protons to neutrons in the Earth affects 
the difference between the charged-current and neutral-current cross-sections as well as
the difference between the cross-sections of neutrinos and anti-neutrinos.
At the energies of interest, however, 
the differences between the cross-sections are not significantly affected by 
the varying ratio of protons to neutrons.  
All-in-all, the local density of the Earth is the largest uncertainty.
It primarily affects the energy for which the ratio of survival probabilities starts to deviate from one. 
For example, for inclined directions of the neutrinos, the path through the Earth is shorter and the average density of the Earth less.
The effect will then start at a higher energy.
The general trend, however, remains.\\[\baselineskip]

It should be noted that the probability of a neutrino to interact in the (vicinity of the) detector and 
to produce a detectable signal is not taken into account.
This requires a convolution with the complete energy spectrum of the neutrinos and a simulation of the detector response.
Nonetheless, the observed increase of the survival probability is interesting.
It could lead to a larger number of detected events from an astrophysical source than previously assumed, 
albeit with some blurring of the image.
This blurring is related to the typical neutrino scattering angle 
which is --at these energies-- usually much smaller than the angular resolution of a neutrino telescope.
It is also interesting to note that the regeneration of tau-neutrinos further contributes to the diffusion and energy loss \cite{ref:halzen-saltzberg}.
As a consequence, the increase of the detectable flux of tau-neutrinos will be larger than that of muon and electron neutrinos.
Note finally that the reported increase {\em partially} applies to a diffuse flux. 
The increase due to the energy loss in neutral-current interactions applies but
the increase from neutrinos outside the target area does not (which can be understood from Liouville's theorem).
Considering that the energy spectrum of neutrinos produced by interactions of cosmic rays in the Earth' atmosphere above the detector 
is generally steeper than that from astrophysical sources,
the expected increase of the number of detectable events should come without serious costs.

\section{Conclusions}

An increase of the detectable flux of neutrinos is found 
in the scenario where charged-current as well as neutral-current interactions are taken into account 
compared to other --commonly used-- scenarios.
As a consequence, 
the estimated number of events from an astrophysical source that could be detected in neutrino telescopes will be larger.

\bibliographystyle{plain}
\bibliography{Earth}

\end{document}